\documentstyle[12pt]{article}
\def\zid{1\kern-0.36em\llap~1}

\newcommand{\beq}{\begin{equation}}
\newcommand{\ber}{\begin{eqnarray}}
\newcommand{\eeq}{\end{equation}}
\newcommand{\eer}{\end{eqnarray}}

\begin{document}

\begin{titlepage}
\rightline{[SUNY BING 4/22/03 ] }
\rightline{ hep-ph/0305036}
\vspace{2mm}
\begin{center}
{\bf \hspace{0.1 cm} CONSEQUENCES OF A LARGE TOP-QUARK CHIRAL
WEAK-MOMENT }\\ \vspace{2mm} Charles A. Nelson\footnote{Electronic
address: cnelson @ binghamton.edu \newline Talk presented at ``Les
Rencontres de la Valle d'Aoste", March 12, 2003. }
\\ {\it Department of Physics, State University of New York at
Binghamton\\ Binghamton, N.Y. 13902}\\[2mm]
\end{center}


\begin{abstract}

This talk concerns some theoretical patterns of the helicity
amplitudes for $t \rightarrow W^{+} b $ decay. The patterns
involve both the standard model's decay helicity amplitudes,
$A_{SM}\left( \lambda _{W^{+} } ,\lambda _b \right)$, and the
amplitudes $A_{+}\left( \lambda _{W^{+} } ,\lambda _b \right)$ in
the case of an additional $t_R \rightarrow b_L$ tensorial coupling
of relative strength $\Lambda_{+} =E_W /2 \sim 53 GeV$. Such an
additional electroweak coupling would arise if the observed
top-quark has a large chiral weak-transition-moment. The $A_{+}$
amplitudes are interpreted as corresponding to the observed
top-quark decays. Three tWb-transformations $A_{+}=M $ $A_{SM},
...$, are used in simple characterization of the values of
$\Lambda_{+}$, $m_W/m_t$, and $m_b/m_t$.  Measurement of the sign
of the $\eta_L = \pm 0.46(SM/+) $ helicity parameter, due to the
large interference between the $W$ longitudinal and transverse
amplitudes, could exclude such a chiral weak-transition-moment in
favor of the SM prediction.
\end{abstract}

\end{titlepage}

\section{Introduction}

While the theoretical analysis discussed in this talk does involve
the observed mass values of the top-quark, $W$ boson, and the
$b$-quark, it is not a matter of any presently available empirical
data disagreeing with a standard model (SM) prediction. Instead,
the interest is because of some theoretical patterns of the
helicity amplitudes for $t \rightarrow W^{+} b $ decay. The
theoretical patterns involve both the standard model's decay
helicity amplitudes, $A_{SM}\left( \lambda _{W^{+} } ,\lambda _b
\right)$, and the amplitudes $A_{+}\left( \lambda _{W^{+} }
,\lambda _b \right)$ in the case of an additional $t_R \rightarrow
b_L$ tensorial coupling of relative strength $\Lambda_{+} =E_W /2
\sim 53 GeV$. To focus the discussion, in this talk the $A_{+}$
amplitudes are interpreted as corresponding to the observed
top-quark decays $t \rightarrow W^{+} b $ [1]. This identification
hypothesis might be excluded by future theoretical analysis and/or
empirical data; in (I), alternatives to this identification were
considered [2]. Experimental tests and measurements in ongoing and
forthcoming [1,3,4] top-quark decay experiments at hadron and $l^-
l^+$ colliders should be able to significantly clarify matters.
The explicit expressions for these amplitudes, and other details,
are given in (I) and in a ``hep-ph" preprint (II) [2].

Measurement of the sign of the $\eta_L \equiv \frac 1\Gamma
|A(-1,-\frac 12)||A(0,- \frac 12)|\cos \beta _L = \pm 0.46(SM/+) $
helicity parameter[5], due to the large interference between the
$W_{Longitudinal}$ and $W_{Transverse}$ amplitudes, could exclude
such a large chiral weak-transition-moment in $t \rightarrow W^{+}
b $ decay in favor of the SM prediction.  On the other hand,
measurement of the SM predicted fraction of final
$W_{Longitudinal}$ versus final $W_{Transverse}$ bosons for this
decay mode would not distinguish between the two cases.  The
definitive empirical test must establish the sign of
$cos(\beta_L)$ where $ \beta_L $ is the relative phase of the two
$\lambda _b = -1/2$ amplitudes, $A \left( 0, - 1/2 \right)$ and $A
\left( -1 , -1/2 \right)$, c.f. Table 1 below.

In the  $t$-quark rest frame, the matrix element for $t
\rightarrow W^{+} b$ is
\begin{equation}
\langle \theta _1^t ,\phi _1^t ,\lambda _{W^{+} } ,\lambda _b
|\frac 12,\lambda _1\rangle =D_{\lambda _1,\mu }^{(1/2)*}(\phi
_1^t ,\theta _1^t ,0)A_{i} \left( \lambda _{W^{+} } ,\lambda _b
\right)
\end{equation}
where $\mu =\lambda _{W^{+} } -\lambda _b $ in terms of the $W^+$
and $b$-quark helicities.  Due to rotational invariance, there are
four independent $A_{i}\left( \lambda _{W^{+} } ,\lambda _b
\right)$ amplitudes for the most general Lorentz coupling. We use
the Jacob-Wick phase-convention for the amplitudes and use the
subscript ``$i$" to identify the amplitude's associated coupling;
in this paper $i=$ SM, $(f_M + f_E)$ for only the additional $t_R
\rightarrow b_L$ tensorial coupling, and $(+)$ for $A_{+}
(\lambda_W, \lambda_b) = A_{SM} (\lambda_W, \lambda_b) + A_{f_M +
f_E } (\lambda_W, \lambda_b)$ when $\Lambda_{+} =E_W /2$. With
respect to the latter case, the Lorentz coupling involving both
the SM's $(V-A)$ coupling and an additional $t_R \rightarrow b_L$
tensorial coupling of arbitrary relative strength $\Lambda_{+}$ is
$ W_\mu ^{*} J_{\bar b t}^\mu = W_\mu ^{*}\bar u_{b}\left(
p\right) \Gamma ^\mu u_t \left( k\right) $ where $k_t =q_W +p_b $,
and
\begin{equation}
\frac{1}{2} \Gamma ^\mu =g_L\gamma ^\mu P_L  + \frac{g_{f_M + f_E}
} {2\Lambda _{+} }\iota \sigma ^{\mu \nu }(k-p)_\nu P_R \\
\end{equation}
Thus, for $\Lambda_{+} = E_W/2$ in $g_L = g_{f_M + f_E} = 1$
units, which corresponds to the (+) amplitudes, the Lorentz
structure of the effective coupling is very simple
\begin{eqnarray}
\gamma ^\mu P_L + \iota \sigma ^{\mu \nu } v_\nu P_R \\
 =P_R \left( \gamma
^\mu + \iota \sigma ^{\mu \nu } v_\nu \right)
\end{eqnarray}
where $P_{L,R} = \frac{1}{2} ( 1 \mp \gamma_5 ) $ and $v_{\nu}$ is
the W-boson's relativistic four-velocity.

The interest in these particular couplings arose as a by-product
of a consideration [6] of future measurements of competing
observables in $t \rightarrow W^+ b $ decay.  In particular, we
considered the SM's the $g_{V-A}$ coupling values of helicity
decay parameters versus those for `` $(V-A)$ $+$ single additional
Lorentz structures." It was found that versus the SM's dominant
L-handed $b$-quark amplitudes, there are two ``dynamical
phase-type ambiguities" produced respectively by an additional
$(S+P)$ coupling and by an additional $t_R \rightarrow b_L$
tensorial coupling, see the $ A \left( 0,-\frac 12\right) $ and $
A \left( -1,-\frac 12\right) $ columns of Table 1. Such a
dynamical-ambiguity produced physically by the additional Lorentz
structure is to be contrasted to the mathematical forcing of a
``phase-ambiguity" by simply changing by-hand the sign of one, or
more, of the four helicity amplitudes $A(\lambda_W, \lambda_b)$.
By tuning the effective-mass-scale associated with the additional
coupling constant, the additional $(S+P)$ coupling, $(f_M + f_E)$
coupling, has respectively changed the sign of the $ A \left(
0,-\frac 12\right) $,  $ A \left( -1,-\frac 12\right) $ amplitude.
In $ g_L = g_{S+P} = g_{+} = 1 $ units, the corresponding
effective-mass scales are $\Lambda_{S + P} \sim -35 GeV$,
$\Lambda_{+} \sim 53 GeV$.  The numerical patterns shown in the
table in the case of the additional $\Lambda_{S + P} $ coupling
are not surprising for the $(S+P)$ coupling because it only
contributes to the $W_{Longitudinal}$ amplitudes. However,
associated with the additional $t_R \rightarrow b_L$ tensorial
coupling, labeled $(f_M + f_E)$ in this table, three interesting
numerical puzzles arise at the $0.1 \%$ level in the (+)
amplitudes versus the SM's pure $(V-A)$ amplitudes.

The 1st puzzle is that the $A_{+} (0,-1/2)$ amplitude has the same
value as the $A_{SM} (-1,-1/2)$ amplitude in the SM; see the
corresponding two ``220" entries in the top of Table 1. From the
empirical t-quark and W-boson mass values, the mass ratio $ y =
\frac{m_W} {m_t} = 0.461 \pm 0.014$. This can be compared with the
puzzle's associated mass relation
\begin{eqnarray}
1-\sqrt{2}y-y^{2}-\sqrt{2}y^{3}=x^{2}(\frac{2%
}{1-y^{2}}-\sqrt{2}y)-x^{4}(\frac{1-3y^2}{(1-y^2)^3})+\ldots
\end{eqnarray}
\begin{eqnarray*}
=1.89x^{2}-0.748x^{4}+\ldots
\end{eqnarray*}
which follows by setting $A_{+} (0,-1/2) = A_{SM}(-1,-1/2) $ and
then expanding in $x^{2}=(m_{b}/m_{t})^{2}$ the $A_{+} (0,-1/2)$
amplitude, with $\Lambda_{+} = E_{W} / 2 = \frac{m_t } {4 } [ 1 +
y^2 -  x^2]$ in $g_L = g_{+} =1$ units. Since empirically
$x^{2}\simeq 7 \cdot 10^{-4}$, there is only a 4th
significant-figure correction from the finite b-quark mass to the
only real-valued solution $y=0.46006$ ($m_{b}=0$) of this mass
relation. The $0.1 \%$ level of agreement of the two ``220"
entries of Table 1 is due to the present central value of $ m_t $,
and to the central value and $ 0.05 \%$ precision of $ m_W$.  The
error in the empirical value of the mass ratio $y$ is dominated by
the current $ 3\%$ precision of $ m_t$.

The 2nd and 3rd numerical puzzles are the occurrence of the same
magnitudes of the two R-handed b-quark amplitudes $A_{New} =
A_{g_L =1} / \sqrt \Gamma $ for the SM and for the additional $t_R
\rightarrow b_L$ tensorial coupling. This is shown in the $ A
\left( 0, \frac 12\right) $ and $ A \left( 1, \frac 12\right) $
columns in the bottom half of Table 1. As explained below, for
$\Lambda_{+} = E_W/2$ the magnitudes of these two R-handed moduli
are actually exactly equal and not merely numerically equal to the
$ 0.1\%$ level.

We will next discuss different types of helicity amplitude
relations involving both the standard model's decay helicity
amplitudes, $A_{SM}\left( \lambda _{W^{+} } ,\lambda _b \right)$,
and the amplitudes $A_{+}\left( \lambda _{W^{+} } ,\lambda _b
\right)$ in the case of an additional $t_R \rightarrow b_L$
tensorial coupling of relative strength $\Lambda_{+}$. These
relations in some cases ``explain" and in other cases analytically
realize as theoretical patterns, these and other numerical puzzles
of Table 1.

Helicity amplitude relations of types (i) and (ii) are exact
ratio-relations holding for all $y = \frac {m_W} {m_t} , x = \frac
{m_b} {m_t} ,$ and $ \Lambda_{+}$ values,. By the type (iii)
ratio-relations holding for all $y = \frac {m_W} {m_t}$ and $ x =
\frac {m_b} {m_t} $ values, the tWb-transformation $A_{+}=M$
$A_{SM}$ where $M=v$ $diag(1,-1,-1,1)$ characterizes the mass
scale $\Lambda_{+} = E_W/2 $. The parameter $v$ is the velocity of
the $W$-boson in the t-quark rest frame.  Somewhat similarly, the
amplitude condition (iv), $A_{+} (0,-1/2) = a A_{SM} (-1,-1/2)$
with $a= 1 + O(v \neq y \sqrt{2}, x)$, and the amplitude condition
(v), $A_{+} (0,-1/2) = - b A_{SM} (1,1/2)$ with $ b = v^{-8} $,
determine respectively the scale of two additional 4x4
tWb-transformation matrices $P$ and $B$. Thereby, (iv) and (v)
characterize the values of the mass ratios $y = m_W/m_t$ and $ x=
m_b/m_t$. $O(v \neq y \sqrt{2}, x)$ denotes small corrections. It
is not understood why the values are simple for the parameters $a$
and $b$.

\section{Helicity amplitude relations}

The first type of ratio-relations holds separately for $i=(SM)$,
$(+)$; (i):
\begin{equation}
\frac{A_{i} (0,1/2) } { A_{i} (-1,-1/2) } = \frac{1}{2}
\frac{A_{i} (1,1/2) } { A_{i} (0,-1/2) }
\end{equation}

The second type of ratio-relations relates the amplitudes in the
two cases  (ii): Two sign-flip relations, note sign changes of
amplitudes in Table 1,
\begin{equation}
\frac{A_{+} (0,1/2) } { A_{+} (-1,-1/2) } = \frac{A_{SM} (0,1/2) }
{ A_{SM} (-1,-1/2) }
\end{equation}
\begin{equation}
\frac{A_{+} (0,1/2) } { A_{+} (-1,-1/2) } = \frac{1}{2}
\frac{A_{SM} (1,1/2) } { A_{SM} (0,-1/2) }
\end{equation}
and two non-sign-flip relations
\begin{equation}
\frac{A_{+} (1,1/2) } { A_{+} (0,-1/2) } = \frac{A_{SM} (1,1/2) }
{ A_{SM} (0,-1/2) }
\end{equation}
\begin{equation}
\frac{A_{+} (1,1/2) } { A_{+} (0,-1/2) } = 2 \frac{A_{SM} (0,1/2)
} { A_{SM} (-1,-1/2) }
\end{equation}

The third type of ratio-relations, follows by determining the
effective mass scale, $\Lambda_{+}$, so that there is an exact
equality for the ratio of left-handed amplitudes (iii):
\begin{equation}
\frac{A_{+} (0,-1/2) } { A_{+} (-1,-1/2) } = -
 \frac{A_{SM} (0,-1/2) } { A_{SM} (-1,-1/2) },
\end{equation}
This was the tuning condition used to produce the dynamical
phase-ambiguities of Table 1 [6].  Equivalently, $ \Lambda_{+} =
E_W/2$ follows from each of:
\begin{equation}
\frac{A_{+} (0,-1/2) } { A_{+} (-1,-1/2) } = -
 \frac{1}{2} \frac{A_{SM} (1,1/2) } { A_{SM} (0,1/2) },
\end{equation}
\begin{equation}
\frac{A_{+} (0,1/2) } { A_{+} (1,1/2) } = - \frac{A_{SM} (0,1/2) }
{ A_{SM} (1,1/2) },
\end{equation}
\begin{equation}
\frac{A_{+} (0,1/2) } { A_{+} (1,1/2) } = - \frac{1}{2}
\frac{A_{SM} (-1,-1/2) } { A_{SM} (0,-1/2) },
\end{equation}
Alternatively, the value of $\Lambda_{+}$ can be characterized by
postulating the existence of a tWb-transformation $A_{+}=M$
$A_{SM}$ where $M=v$ $diag(1,-1,-1,1)$, with \newline $
A_{SM}=[A_{SM}(0,-1/2),A_{SM}(-1,-1/2),A_{SM}(0,1/2),A_{SM}(1,1/2)]$
and analogously for $A_{+}$.

Assuming (iii), the fourth type of relation is the equality (iv):
\begin{equation}
A_{+} (0,-1/2) = a A_{SM} (-1,-1/2),
\end{equation}
where $a= 1 + O(v \neq y \sqrt{2}, x)$. This is equivalent to the
velocity formula $  v = a  y \sqrt{2} \left( \frac{1} {1- ( E_b -
q )/m_t} \right)
 \simeq a y \sqrt{2} ,$ for $ m_b = 0 $.
For $a=1$, (iv) leads to the mass relation discussed above,
Eq.(5).  However, for $a=1$, (iv) also leads to $\sqrt{2}=v\gamma
(1+v)=v\sqrt{\frac{1+v}{1-v}}$ so $v=0.6506\ldots$ without input
of a specific value for $m_b$. But by Lorentz invariance $v$ must
depend on $m_b$. Accepting (iii) as exact, we interpret this to
mean that $a \neq 1$.  As shown in (II), the $O(v \neq y \sqrt{2},
x)$ corrections in $a$, required by Lorentz invariance, arise from
$v \neq y \sqrt{2}$ and $x \neq 0$.

Equivalently, for $a$ arbitrary, (15) can be expressed postulating
the existence of a second tWb-transformation $A_{+}=P $ $A_{SM}$
where
\begin{equation}
P\equiv v \left[
\begin{array}{cccc}
0 & a/v & 0 & 0 \\ -v/a & 0 & 0 & 0 \\ 0 & 0 & 0 & -v/2a \\ 0 & 0
& 2a/v & 0
\end{array}
\right]
\end{equation}

The above two tWb-transformations do not relate the $\lambda_b= -
\frac{1}{2}$ amplitudes with the $\lambda_b= \frac{1}{2}$
amplitudes. From (i) thru (iv), in terms of a parameter $b$, the
equality (v):
\begin{equation}
A_{+} (0,-1/2) = - b A_{SM} (1,1/2),
\end{equation}
is equivalent to $A_{+}=B $ $A_{SM}$
\begin{equation}
B\equiv \left[
\begin{array}{cccc}
0 & 0 & 0 & -b \\ 0 & 0 & 2b & 0 \\ 0 & v^{2}/2b & 0 & 0
\\ -v^{2}/b & 0 & 0 & 0
\end{array}
\right]
\end{equation}
The choice of $ b = v^{-8} = 31.152$, gives
\begin{equation} B\equiv v
\left[
\begin{array}{cccc}
0 & 0 & 0 & -v^{-9} \\ 0 & 0 & 2v^{-9} & 0 \\ 0 & v^{9}/2 & 0 & 0
\\ -v^{9} & 0 & 0 & 0
\end{array}
\right]
\end{equation}
and corresponds to the mass relation $ m_{b}
=\frac{m_{t}}{b}\left[ 1-\frac{vy}{\sqrt{2}}\right] =4.407...GeV $
for $m_t = 174.3 GeV$.

If one does not distinguish the $(+)$ versus SM indices,
respectively of the rows and columns, these three
tWb-transformation matrices have some simple properties, for
details see (II): The anticommuting 4x4 matrices $M, P$ with $a$
arbitrary, and $Q$ satisfy the closed algebra
$[\overline{M},\overline{P}]=2\overline{Q},[\overline{M},\overline{Q}]
=2\overline{P},[\overline{P},\overline{Q}]=2\overline{M}$. The bar
denotes removal of the overall ``$v$" factor, $M= v \overline{M},
...$.  Note that $Q$ is not a tWb-transformation; $Q$ is obtained
from the first listed commutator.

Including the B matrix with both $a$ and $b$ arbitrary, the
``commutator + anticommutator" algebra closes with 3 additional
matrices $C, H,
G$ obtained by $\{\overline{%
M},\overline{B}\}=-2\overline{C};
[\overline{P},\overline{B}]=2\overline{H}$; and $\{\overline{P},\overline{%
C}\}=-2\overline{G}$. This has generated an additional tWb transformation $G\equiv v\overline{G}$%
; but $C\equiv v\overline{C}$ and $H\equiv v\overline{H}$ are not
tWb transformations.

\section{Discussion}

The elements of the three logically-successive tWb transformations
are constrained by the exact helicity amplitude ratio-relations
(i) and (ii). Thereby, the type (iii) ratio-relation fixes
$\Lambda_{+} = E_W/2$ and the overall scale of the
tWb-transformation matrix $M$. Somewhat similarly, the amplitude
condition (iv) with $a= 1 + O(v \neq y \sqrt{2}, x)$ and the
amplitude condition (v) with $ b = v^{-8} $ determine respectively
the scale of the tWb-transformation matrices $P$ and $B$ and
characterize the values of $m_W/m_t$ and $m_b/m_t$. The overall
scale can be set here by $ m_t $ or $ {m_W}$. From the perspective
of further ``unification", $ m_W$ is more appropriate since its
value is fixed in the SM.

The additional $t_R \rightarrow b_L$ tensorial coupling violates
the conventional gauge invariance transformations of the SM and
traditionally in electroweak studies such anomalous couplings have
been best considered as ``induced" or ``effective". The $f_E$
component corresponds to a ``second class current" [7]. $f_E$ has
a distinctively different reality structure, and time-reversal
invariance property versus the first class $ V, A, f_M$ [8].

In the present context, supersymmetry could provide a more general
and useful off-shell theoretical framework in which to consider
these theoretical patterns of the helicity amplitudes for $t
\rightarrow W^{+} b $ decay. Form factor effects would naturally
occur. In the extant MSSM literature, see more complete references
in (II), sizable ``one-loop-level" reductions in the $t
\rightarrow W^{+} b $ partial decay width have been reported: From
SM Higgs and additional MSSM's Higg's there is a small $ \leq 2 \%
$ correction. However, from SUSY electroweak corrections, there is
in [9] an up to $ 10\% $ reduction, depending on $tan ( \beta ) $.
From QCD including some two-loop-level corrections and SUSY QCD
corrections in the summary of [10] a $ 25\% $ reduction is
reported. It is to be emphasized that, firstly, the (+) partial
width considered in the present paper constitutes a very large,
net $ 56 \% $ reduction versus the Born-level SM value and that,
secondly, these cited SUSY calculations have been for the partial
width, so other couplings instead of an additional effective $t_R
\rightarrow b_L$ tensorial coupling, might be predominantly
responsible for these reported reductions.

\section{Experimental Tests/Measurements}

Empirically, important tests of the physical relevance of the
theoretical patterns to the observed top-quark decays are:
\newline (a) Measurement of the sign of the $\eta_L \equiv \frac
1\Gamma |A(-1,-\frac 12)||A(0,- \frac 12)|\cos \beta _L = \pm
0.46(SM/+) $ helicity parameter via determination of stage-two
spin-correlation observables [5] for the
$t\overline{t}\longrightarrow l\overline{l}+jets$ channel.  These
values for $\eta_L$ are essentially the maximal possible
deviations since $ | \eta_L | = 0.5$ is the kinematic limit.  The
differences from $ | \eta_L | = 0.5$ are due to $m_b \neq 0$.
 \newline
 (b) Measurement of the
closely associated $ {\eta_L}^{'} \equiv \frac 1\Gamma
|A(-1,-\frac 12)||A(0,- \frac 12)|\sin \beta _L $ helicity
parameter.  This would provide useful complementary information,
since in the absence of $T_{FS}$-violation, ${\eta_L}^{'} =0$ [6].
$T_{FS}$-violation can occur due to intrinsic time-reversal
violation and/or large $W^+ b$ final-state interactions.  It is
very important to exclude sizable $T_{FS}$-violation and/or
$CP$-violation in top-quark decays.
\newline
 (c) Measurement of the partial width for $t \rightarrow
W^+ b$, e.g. by single top-quark production at a hadron collider
[11]. The $v^2$ factor which differs their associated partial
widths corresponds to the SM's $\Gamma_{SM}= 1.55 GeV$, versus
$\Gamma_+ = 0.66 GeV$ and a longer-lived (+) top-quark if this
mode is dominant.

{\bf Acknowledgments: }

We thank experimental and theoretical physicists for discussions,
and the Fermilab Theory Group for a visit during the summer of
2002. This work was partially supported by U.S. Dept. of Energy
Contract No. DE-FG 02-86ER40291.

\begin{center}
{\bf Table Captions}
\end{center}

Table 1:  Numerical values of the helicity amplitudes $A \left(
\lambda _{W^{+} } ,\lambda _b \right)$ for the standard model and
for the two dynamical phase-type ambiguities (with respect to the
SM's dominant $\lambda_b= - \frac{1}{2}$ amplitudes).  The values
are listed first in $ g_L = g_{f_M + f_E} = 1 $ units, and second
as $ A_{new} = A_{g_L = 1} / \surd \Gamma $ where $\Gamma$ is the
partial width for $t \rightarrow W^{+} b $. [ $m_t=175GeV, \; m_W
= 80.35GeV, \; m_b = 4.5GeV$ ].

\end{document}